\documentclass[aps,prb,twocolumn]{revtex4-1}
\usepackage{graphicx}
\usepackage{amsmath}
\usepackage{amsfonts}
\usepackage{color}
\usepackage{mathtools}
\usepackage{braket}
\usepackage{gensymb}
\usepackage{textcomp}
\usepackage{soul}
\usepackage[colorlinks=true,linkcolor=blue,citecolor=blue,urlcolor=blue]{hyperref}

\begin{document}

\title{Pressure-induced magnetic and topological transitions in non-centrosymmetric MnIn$_{2}$Te$_{4}$}
\author{Anumita Bose}
\email{anumitabose@iisc.ac.in}
\affiliation{Solid State and Structural Chemistry Unit, Indian Institute of Science, Bangalore 560012, India}

\author{Rajdeep Banerjee}
\email{rajdeep.jzs@gmail.com}
\affiliation{Solid State and Structural Chemistry Unit, Indian Institute of Science, Bangalore 560012, India}

\author{Awadhesh Narayan}
\email{awadhesh@iisc.ac.in}
\affiliation{Solid State and Structural Chemistry Unit, Indian Institute of Science, Bangalore 560012, India}

\date{\today}

\begin{abstract}
The discovery of time-reversal-invariant topological states has drawn great attention in recent decades. However, despite the potential of displaying a variety of exotic physics, the study of magnetic topological phases lags behind due to underlying added complexity of magnetism. In this work, we predict the interplay of magnetism and topology in the non-centrosymmetric ternary manganese compound MnIn$_2$Te$_4$, using first-principles calculations. At ambient pressure, the ground state of the system is an antiferromagnetic insulator. With the application of small hydrostatic pressure ($\sim$0.50 GPa), it undergoes a magnetic transition and the ferromagnetic state becomes energetically favourable. At $\sim$2.92 GPa, the system undergoes a transition into a Weyl semimetallic phase, which hosts multiple Weyl points in the bulk and is associated with non-trivial surface Fermi arcs. Remarkably, we discover that the number of Weyl points in this system can be controlled by pressure and that these manifest in an anomalous Hall conductivity (AHC). In addition to proposing a new candidate magnetic topological material, our work demonstrates that pressure can be an effective way to induce and control topological phases, as well as AHC, in magnetic materials. These properties may allow our proposed material to be used as a novel pressure-controlled Hall switch.
\end{abstract}

\maketitle

\textcolor{blue}{\textit{Introduction: }}In recent times, topological phases of matter have attracted considerable attention in condensed matter physics and materials science. The study of time-reversal-invariant topological states in quantum materials has made tremendous progress and created fruitful outcomes in both the theoretical and experimental fields of research~\cite{hasan2010colloquium,hasan2011three,qi2011topological,ando2013topological,armitage2018weyl,burkov2016topological,gao2019topological,weng2016topological_1,kou2017two}. 

The interplay between magnetism and topology further leads to exotic phases, such as quantum anomalous Hall effect (QAHE)~\cite{he2014quantum,liu2016quantum,nomura2011surface,chang2013experimental}, topological magneto-electric effect~\cite{nomura2011surface}, topological axion insulators~\cite{li2010dynamical}, and topological superconductors~\cite{qi2010chiral}. However, due to the added complexity of magnetism, the study of magnetic topological phases has not progressed as rapidly. Traditionally, two methods have been used to unite non-trivial topology and magnetism: (i) doping magnetic elements into systems hosting non trivial topological phases~\cite{chang2013experimental,liu2017dimensional,xu2020high}, and (ii) constructing hetero-structures out of the two~\cite{lang2014proximity,mogi2017tailoring,yasuda2016geometric}. However, in both cases, systems are usually far from satisfactory due to inhomogeneity induced in the doped systems and inhomogeneous band bending near the interface in hetero-structures. Also, since the coupling between topological and magnetic states is usually weak and sensitive to the interface properties, experimental realization of robust systems is quite challenging. Thus, it has become crucial to find magnetic topological phases which are either intrinsic or can be obtained via small perturbations.

In recent years, the study of MnBi$_2$Te$_4$ family of materials have provided a promising platform to explore the combined effect of magnetism and topology intrinsically~\cite{otrokov2019prediction,li2019intrinsic}. These van der Waals materials show a wide variety of topological states depending on the number of layers~\cite{zhang2019topological,otrokov2019unique}. Several other intrinsic magnetic topological phases have been theoretically predicted in the last decade~\cite{xu2011chern,wan2011topological,tang2016dirac,hua2018dirac}. Time reversal symmetry broken Weyl points have been recently studied both theoretically and experimentally in centrosymmetric crystals, such as Fe$_3$GeTe$_2$~\cite{kim2018large}, Co$_3$Sn$_2$S$_2$~\cite{wang2018large,liu2018giant} and Co$_2$MnGa~\cite{belopolski2019discovery}. Importantly, these materials exhibit a giant anomalous Hall effect (AHE) as an outcome of the Berry curvature around the Weyl nodes.

\begin{figure}
\includegraphics[scale=0.22]{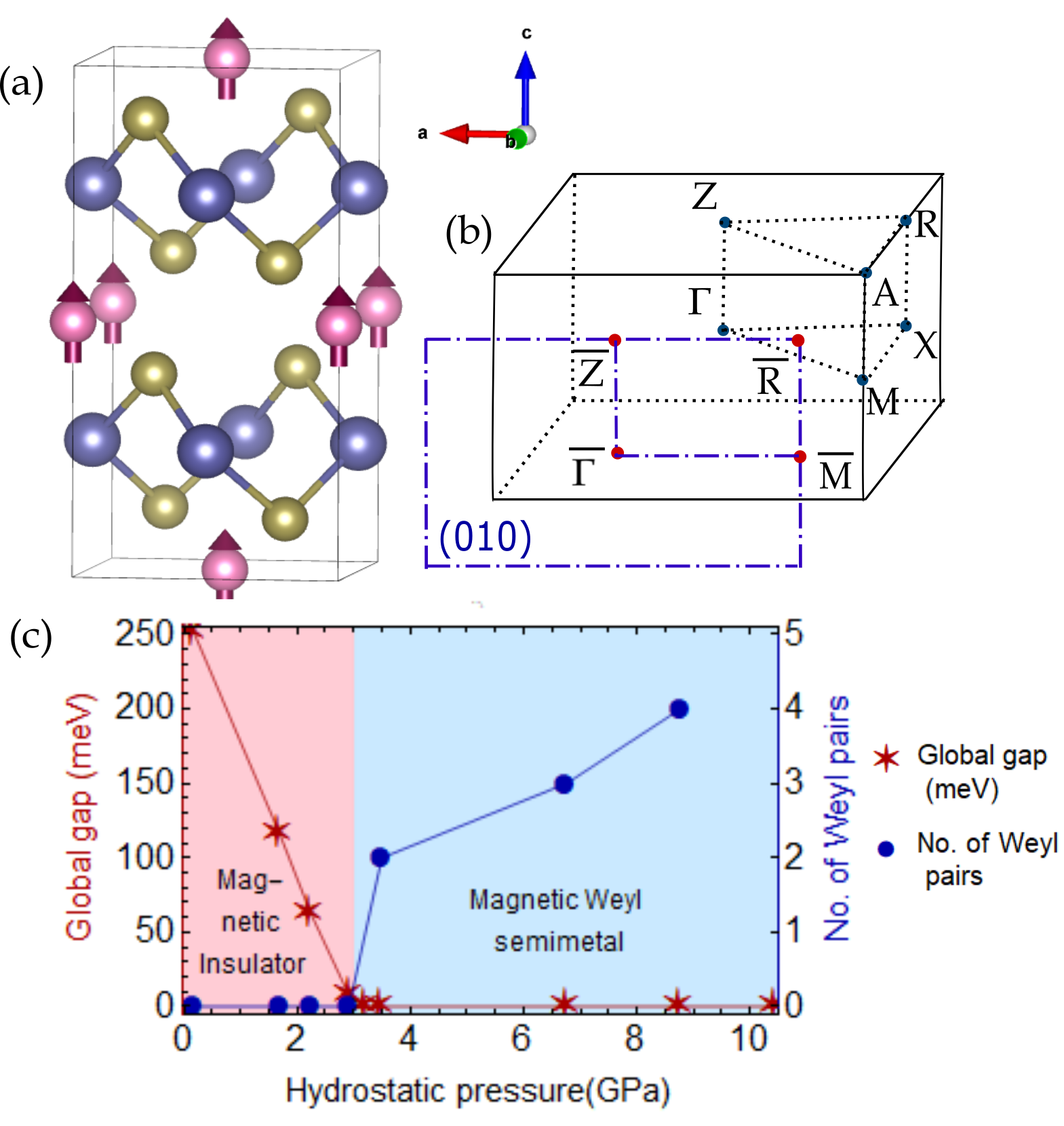}
  \caption{\textbf{Structure and phase diagram of MnIn$_{2}$Te$_{4}$.} (a) Tetragonal unit cell of MnIn$_{2}$Te$_{4}$ with space group $I\Bar{4}2m$ (number 121), comprising of two Mn, four In and eight Te atoms which are denoted by pink, blue and yellow spheres, respectively. The stacking of atomic layers is along the $c$ direction. FM arrangement of Mn spins is depicted as red arrows. (b) Brillouin zone with the high symmetry points and paths marked as blue dots and lines respectively. Rectangular surface with blue dash-dotted line denotes the (010) surface with projection of bulk high symmetry points marked as red dots. (c) The phase transitions with pressure in MnIn$_2$Te$_4$ between magnetic insulator (shaded with light red) and magnetic Weyl semimetal (shaded with light blue). Red line shows the variation of gap between the top-most valence band and bottom-most conduction band, as a function of hydrostatic pressure, which gradually decreases with increasing pressure. Blue line indicates the number of Weyl pairs formed between those two bands in the magnetic Weyl semimetallic phase.}
  \label{unitcell_BZ}
\end{figure}

\begin{figure*}
\includegraphics[scale=0.24]{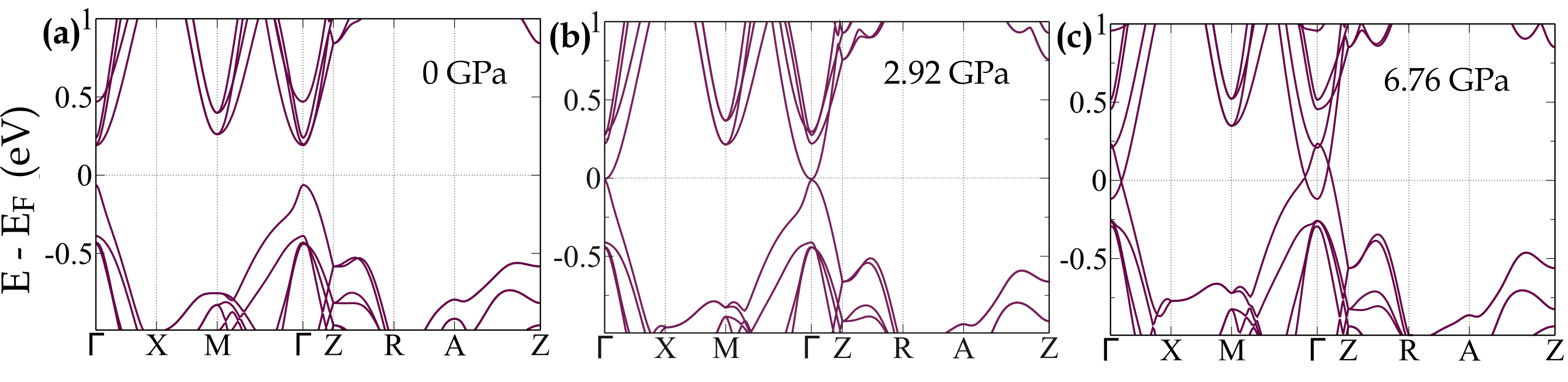}
  \caption{\textbf{Bulk electronic structure of ferromagnetic MnIn$_{2}$Te$_{4}$.} (a) The zero pressure band structure of ferromagnetic MnIn$_{2}$Te$_{4}$, which is a trivial insulator with a direct band gap of nearly 250 meV at $\Gamma$. With increasing hydrostatic pressure, the gap decreases. At 2.92 GPa, VBM and CBM touch each other and the gap becomes zero, which is shown in (b). On increasing the hydrostatic pressure further, overlap between the bands at the Fermi energy increases, which leads to the FM Weyl semimetallic phase. (c) The FM Weyl semimetallic phase at 6.76 GPa, where nodes appear near the $\Gamma$ point close to the Fermi energy.}
  \label{qe_bands}
\end{figure*}

Furthermore, external perturbations such as temperature~\cite{monserrat2016temperature,antonius2016temperature,monserrat2019unraveling}, hydrostatic pressure~\cite{kirshenbaum2013pressure,he2016pressure,zhang2017evidence,sadhukhan2022pressure,juneja2018pressure}, strain~\cite{liu2014tuning,zeljkovic2015strain,narayan2015class,bose2021strain} and light~\cite{rodriguez2021light} allow manipulating topological ordering and its underlying mechanisms. Among these stimuli, the application of hydrostatic pressure is particularly of great utility, as it allows studying the interplay of charge, spin, orbital and lattice degrees of freedom under controlled conditions~\cite{mao2018solids}. 

In this work, we report the first-principles prediction of topological properties of non-centrosymmetric magnetic ternary compound MnIn$_{2}$Te$_{4}$. We show that the ground state of MnIn$_{2}$Te$_{4}$ is an antiferromagetic (AFM) insulator and the material undergoes a magnetic phase transition to a ferromagnetic (FM) insulating phase at $\sim$0.5 GPa. Further increase in hydrostatic pressure results in reduction of the band gap and the emergence of a semimetallic phase with multiple pairs of Weyl points in the system. We confirm the presence of Weyl points by calculating the resulting non-trivial Fermi arcs. Our results also reveal a tunable material system where the number of Weyl points can be controlled by hydrostatic pressure -- these manifest in a non-zero anomalous Hall conductivity (AHC) around the Fermi level. In a nutshell, we propose a new member in the catalogue of magnetic topological materials where both the magnetic ordering (AFM versus FM) and the topology can be controlled by externally applied hydrostatic pressure. The system shows a pressure dependent AHC, which makes the system a potential candidate for designing a pressure-controlled Hall switch. \\

\textcolor{blue}{\textit{Methods: }} Our first-principles calculations were performed within the density functional theory framework as implemented in the Quantum Espresso package~\cite{giannozzi2009quantum,giannozzi2017advanced}, using the Perdew-Burke-Ernzerhof exchange-correlation functional~\cite{perdew1996generalized} and fully relativistic ultra-soft pseudopotentials~\cite{vanderbilt1990soft}. Semi-empirical Grimme’s DFT-D2 correction~\cite{grimme2006semiempirical} was also included to take into account the van der Waals forces. Self-consistent calculations for the bulk were performed using a plane wave cutoff of $50$ Ry and $8\times8\times4$ $\Gamma$-centered Monkhorst-Pack $k$-point grid \cite{monkhorst1976special}. We used an onsite Hubbard $U$ correction for Mn $d$ orbitals, ($U = 5.34$ eV)~\cite{liechtenstein1995density}. During structural optimization the system was relaxed until Hellmann-Feynmann forces on the atoms became less than $10^{-3}$ Ry/bohr. In order to create hydrostatic pressure on the system in our simulations, we applied a chosen strain along the $a$, $b$ and $c$ directions, allowing the atoms to fully relax. The structures for which the diagonal elements of the stress tensor become equal were chosen for the particular hydrostatic pressure value. We further constructed tight-binding models obtained from maximally localized Wannier functions using the {\sc wannier90} code~\cite{mostofi2008wannier90}, with Mn 3$d$, In 5$p$ and Te 5$p$ orbitals as the basis. In order to calculate the position of the bulk Weyl nodes, we used the Nelder and Mead’s Downhill Simplex Method~\cite{nelder1965mead}, as implemented in WannierTools~\cite{wu2018wanniertools}, which takes a uniform $k$-mesh in the three-dimensional Brillouin zone as a set of starting points for the calculation. After obtaining the positions of gapless points, a small sphere is constructed around each and the enclosed Weyl point's chirality is determined by tracking the evolution of the sum of hybrid Wannier charge centers as a function of the polar angle. Surface Green's function method for semi-infinite systems~\cite{guinea1983effective,sancho1984quick,sancho1985highly}, implemented in WannierTools package was employed to calculate the surface electronic spectra. \\

$\textcolor{blue}{\textit{Structural properties:} }$ MnIn$_{2}$Te$_{4}$ is an existing compound which was first synthesized in 1974~\cite{range1975kristallstruktur}. This ternary manganese compound  crystallizes in a non-centrosymmetric tetragonal space group I$\Bar{4}2m$ (number 121). The unit cell has two Mn atoms, four In atoms and eight Te atoms, which are represented as pink, blue and yellow spheres, respectively, in Fig.\,\ref{unitcell_BZ} (a). Due to the presence of Mn spins, the system shows a magnetic nature with broken time reversal symmetry (TRS). Our calculated average magnetic moment per Mn atom is very close to 5 ${\mu}_{B}$  [denoted as red arrows in Fig.\,\ref{unitcell_BZ} (a)], suggesting a half-filed $d$-orbital manifold. At ambient conditions (zero applied pressure), the ground state of the system is found to be an AFM. Although application of hydrostatic pressure makes the FM state energetically more stable compared to the AFM one, as we shall see. The cell parameters for both the FM and AFM systems at ambient pressure obtained from our $\textit{ab-initio}$ calculations are listed in TABLE~\ref{table:unitcell} along with the experimentally measured values~\cite{range1975kristallstruktur,doll1991structural}. We find that the optimized lattice parameters match well with the experimental values. At ambient pressure conditions, AFM phase is energetically more stable (E$_{AFM}$-E$_{FM}$ = 7.16 meV). However, at small hydrostatic pressure values of $\sim$0.5 GPa, the FM state becomes energetically more favourable. Next, we will look at the pressure variation of the electronic structure. 

\begin{table}
\caption{\textbf{Comparison of cell parameters for different magnetic orders.} The calculated cell parameters obtained from first principles calculations for FM and AFM magnetic orders. Experimental lattice parameters are from Refs.~\cite{range1975kristallstruktur,doll1991structural}.}
\label{table:unitcell}
\begin{tabular}{p{2.5cm} p{1.8cm} p{1.8cm}}
\hline
\hline
 Magnetic order & $a=b$ (\AA{}) & $c$ (\AA{})\\
 \hline
 \hline
  AFM  & 6.138 & 12.229\\
  FM & 6.172 & 12.324\\
  Experimental &6.191& 12.382\\
 \hline
 \hline
\end{tabular}
\end{table}

$\textcolor{blue}{\textit{Bulk electronic structure: }}$ In Fig.\,\ref{qe_bands}, we examine the bulk electronic band structures of the FM MnIn$_{2}$Te$_{4}$ system, at different values of hydrostatic pressure, including spin orbit coupling (SOC), along the high symmetry directions. At zero pressure, the system is a trivial magnetic insulator [shown in Fig.\,\ref{qe_bands} (a)] with a direct band gap of nearly 250 meV at the $\Gamma$ point. The top most valence band has mainly Te $p$ orbital character, while Te $p$ and In $s$ orbitals contribute most to the lowest conduction band. For a comparison of the band structure of the AFM phase, see supplemental material~\cite{abc}. With increasing hydrostatic pressure, the gap between valence band maximum (VBM) and conduction band minimum (CBM) shrinks continuously. Variation of the direct band gap as a function of hydrostatic pressure can be seen in Fig.\,\ref{unitcell_BZ} (c). The band gap decreases almost linearly with increasing hydrostatic pressure. Here, the region shaded with light red shows the gapped phase. As the hydrostatic pressure is further increased, the valence and conduction band edges meet at the Fermi energy at the $\Gamma$ point, where the gap vanishes and we observe a phase transition to a semimetallic phase at $\sim$2.92 GPa, which can be seen from Fig.\,\ref{qe_bands} (b). Fig.\,\ref{qe_bands} (c) shows the bulk bandstructure of the semimetallic phase at a representative pressure value of 6.76 GPa. The bands around $\Gamma$ near the Fermi energy get inverted, creating gapless points around the center of the BZ. Although, we note that, the nodes appear away from the high symmetric directions. For a list of the position of Weyl nodes in the three-dimensional momentum space along with their respective chirality values, see supplemental material~\cite{abc}. The blue dots in Fig.\,\ref{unitcell_BZ} (c) indicate the number of Weyl point pairs present in the semi-metallic phase near the Fermi energy. Remarkably, we find that the number of Weyl points monotonically increases as one varies the hydrostatic pressure in the system from 2.92 GPa to 8.77 GPa. Thus, hydrostatic pressure not only creates Weyl points in our system, but also allows control over the number of Weyl points. This intriguing feature may be directly probed in near-future experiments. We further note that being non-centrosymmetric, MnIn$_2$Te$_4$ lacks inversion symmetry. On the other hand, TRS is also broken in the system due to its magnetic nature. This means that the Weyl points in this system occur in absence of both the inversion and time reversal symmetry, indicating the robustness of the these points to structural distortions. \\

\begin{figure}
\includegraphics[scale=0.22]{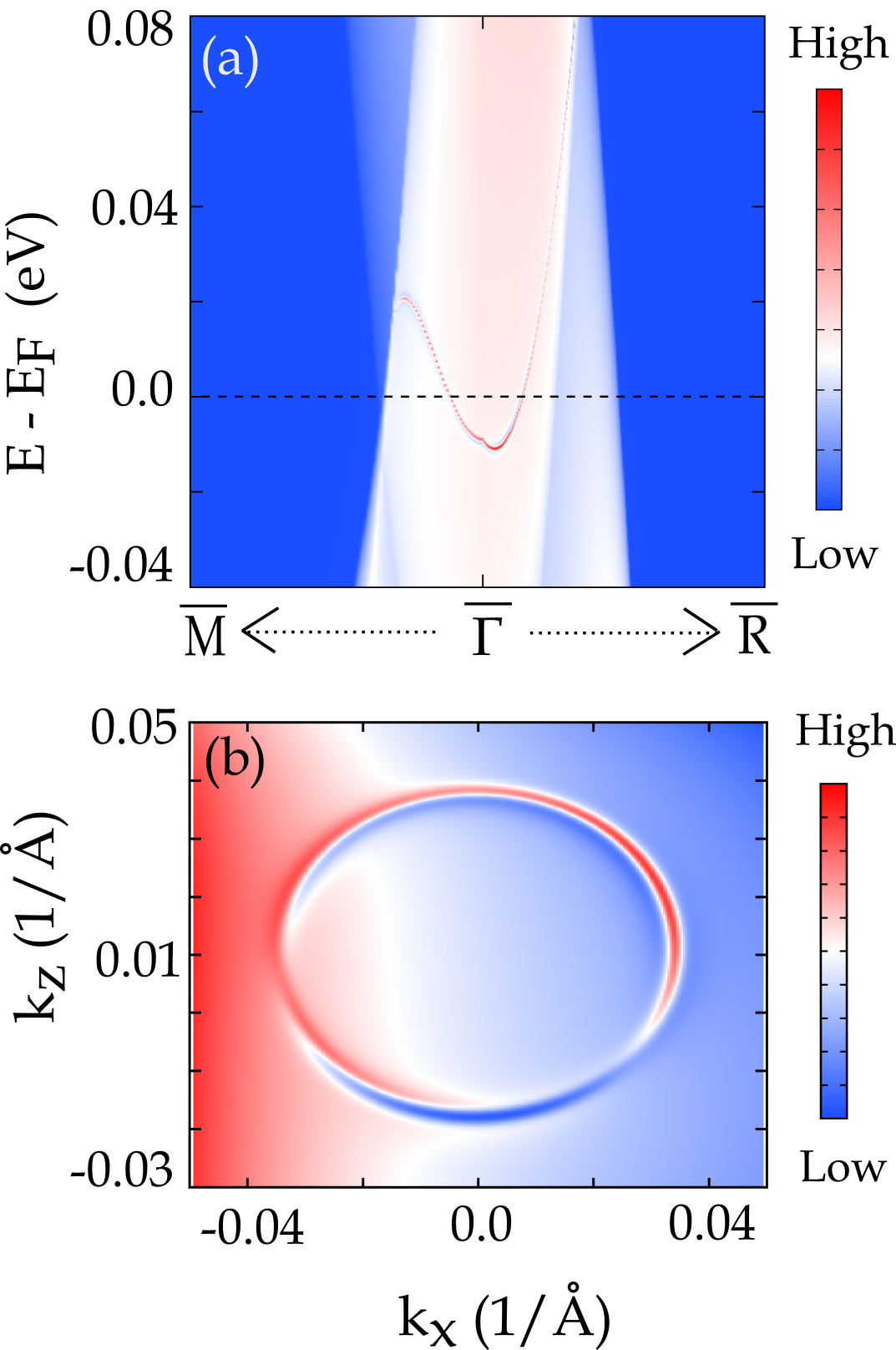}
  \caption{\textbf{Surface electronic spectra.} (a) The spectrum on the (010) surface for a hydrostatic pressure value 6.76 GPa. Existence of surface states is observed along $\Bar{M}-\Bar{\Gamma}-\Bar{R}$ direction. (b) The Fermi arc as a function of the in-plane momenta $k_x$ and $k_z$ at the Fermi level [denoted with the black dashed line in (a)].}
  \label{surface_010}
\end{figure}

\textcolor{blue}{\textit{Surface properties: }}The existence of non-trivial states on the surface is one of the most significant consequences of the occurrence of bulk Weyl points in a three-dimensional system. In order to further demonstrate non-trivial bulk topology of MnIn$_2$Te$_4$, we explore the surface electronic spectra of the system on the (010) surface. On this surface, bulk high symmetry points M, $\Gamma$ and Z project on $\Bar{M}$, $\Bar{\Gamma}$ and $\Bar{Z}$, respectively, as shown in Fig.\,\ref{unitcell_BZ} (b). Fig.\,\ref{surface_010} (a) shows the (010) surface state along $\Bar{M}-\Bar{\Gamma}-\Bar{R}$ direction connecting the bulk Weyl cones extended in the energy window of -0.015 eV to 0.08 eV, for the representative pressure value of 6.76 GPa. With the Weyl points with opposite chirality being energetically separated, we expect to observe Fermi arcs for a range of energy values. Surface states as a function of in-plane momenta $k_x$ and $k_z$, taken at constant energy equal to the Fermi energy $E = E_F$ [shown by the dashed line in Fig.\,\ref{surface_010} (a)], is shown in Fig.\,\ref{surface_010} (b). Here the arcs, connecting the Weyl points with opposite chirality, take the shape of an ellipse. As one goes towards higher energies, the size of the ellipse increases, which can also be observed from Fig.\,\ref{surface_010} (a). We have also studied the surface spectra for other surfaces, for e.g. (100) and (110), and obtained non-trivial surface states. \\

\begin{figure}[h!]
\includegraphics[scale=0.18]{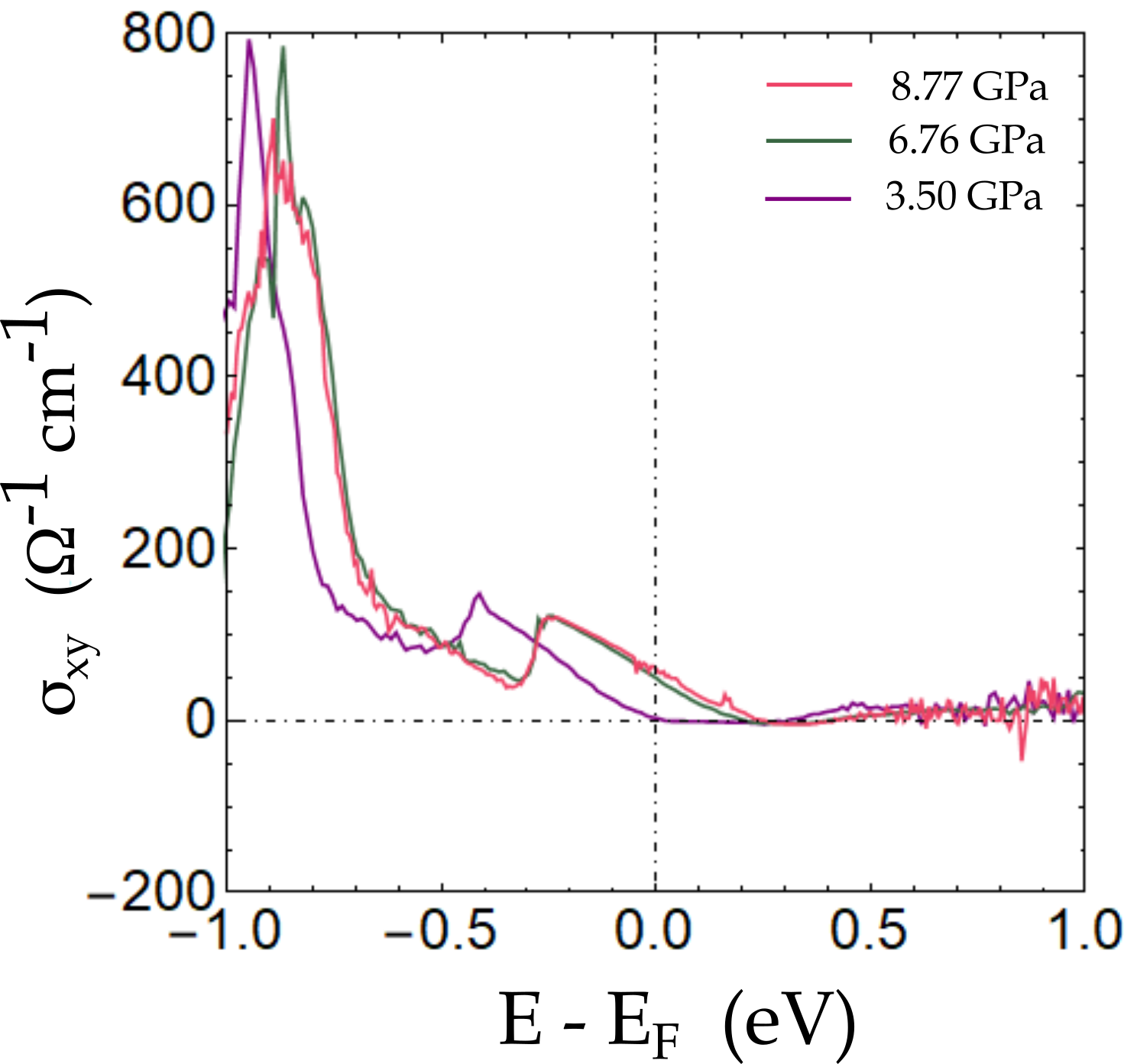}
  \caption{\textbf{Pressure tuned anomalous Hall conductivity.} Violet, green and red curves denote the AHC ($\sigma_{xy}$) for hydrostatic pressure values 3.50 GPa, 6.76 GPa and 8.77 GPa, respectively.  For 3.50 GPa, $\sigma_{xy}$ at the Fermi level is small but non-zero. At 6.76 GPa, it experiences a sudden jump, which is saturated for higher pressure values. With increasing pressure, the number of Weyl points increases, leading to an increase in $\sigma_{xy}$ at Fermi level with pressure.}
  \label{AHC_pressure}
\end{figure}

\textcolor{blue}{\textit{Anomalous Hall effect: }} Next, we explore another intriguing aspect of our system, namely the anomalous Hall effect. Recent studies have established that intrinsic AHC, $\sigma_{xy}$, which originates from a nonzero Berry curvature, can be induced in systems with Weyl points~\cite{fang2003anomalous,burkov2014anomalous,nayak2016large,manna2018colossal}. The AHC is defined as the integral of the Berry curvature of Bloch wave functions over the entire BZ. For the $n$-th band with $z$ component of the Berry curvature $\Omega_{n}^{z}(\textbf{k})$, it reads

\begin{equation}
 \sigma_{xy} = \frac{e^2}{\hbar} \sum_{n} \int_{BZ}^{} f(\epsilon_{n}(\textbf{k})) \Omega_{n}^{z}(\textbf{k}) \frac{d\textbf{k}}{(2\pi)^3},   
\end{equation}
 
where $f(\epsilon_{n}(\textbf{k}))$ denotes the Fermi-Dirac distribution. We next study the AHC for the Weyl phase of our predicted material. \\

Fig.\,\ref{AHC_pressure} shows the variation of AHC component $\sigma_{xy}$ with energy for different values of the applied hydrostatic pressure. We can observe, for 3.50 GPa, $\sigma_{xy}$ is quite small near the Fermi energy, although it increases for 6.76 GPa and gets saturated for higher pressure values. From Fig.~\ref{unitcell_BZ} (c) we can see that the number of Weyl points increases almost linearly with the increase in hydrostatic pressure. This observation may encourage us to expect a linear increase in the AHC, as opposed to the observed saturation. To explain this phenomenon, instead of looking at the total number of Weyl points, we focus only on the Weyl points very close to the Fermi energy. Here, to include the nearest Weyl point energy at 3.50 GPa, we take the window of $|E-E_{F}| \leq$ 0.015 eV, so that at 3.50 GPa we have one Weyl point `close' to the Fermi energy. If this Weyl point is responsible for the small AHC at 3.50 GPa, then we expect the number of Weyl points close to the Fermi energy to increase when the pressure is increased from 3.50 GPa to 6.76 GPa. We see that indeed the number of Weyl points close to Fermi energy, jumps from 1 to 5. A further increase in the pressure from 6.76 GPa to 8.77 GPa does not have much effect on the AHC, and the number of Weyl points close to the Fermi energy also remain fixed at 5. We note that the system becomes metallic at 10.43 GPa, and there are 2 Weyl points close to the Fermi energy at that pressure. The above observations indicate that the number Weyl points very close to the Fermi energy are the ones that control the behaviour of AHC in the vicinity of the Fermi level.

This property enables us to propose this system as a pressure-controlled Hall switch. In an electrical switch a voltage controls the flow of current through a circuit, and beyond a certain voltage, one obtains a current flow. In an analogous manner, here, the pressure controls the AHC at the Fermi energy. Up to 3.50 GPa pressure, the AHC is close to zero. Beyond a certain pressure (6.76 GPa), it becomes significant ($\sim$51 $\ohm^{-1} cm^{-1} $) and maintains the AHC value up to 8.77 GPa. The breakdown pressure of the switch is 10.43 GPa, where the system becomes metallic. \\

\textcolor{blue}{\textit{Summary and conclusions: }}In summary, we have introduced a non-centrosymmetric magnetic material MnIn$_{2}$Te$_{4}$ as a platform to explore the interplay of magnetism and topology. The system is an AFM at ambient pressure conditions and undergoes a magnetic transition to the FM phase at a pressure value $\sim$0.50 GPa. Increase in hydrostatic pressure in FM MnIn$_{2}$Te$_{4}$ further induces a transition from a trivial magnetic insulator to a magnetic Weyl semimetal at $\sim$2.92 GPa, which is experimentally easily achievable. Remarkably, we have shown that the number of Weyl pairs can be tuned by the applied hydrostatic pressure. The existence of these robust Weyl points is confirmed by the presence of non-trivial surface state near the Fermi level and Fermi arcs. As a result of the occurrence of Weyl points, the non-trivial magnetic semi-metallic phase gives rise to a pressure dependent non zero AHC at the Fermi level in the system. This property paves way for its possible application as a pressure-controlled Hall switch. Our findings demonstrate that pressure is a useful tool to unite magnetism with topology and hence produce non-trivial magnetic topological phases. \\

\textcolor{blue}{\textit{Acknowledgements: }}AB acknowledges Prime Minister's Research Fellowship for financial support. RB and AN acknowledge support from the start-up grant (SG/MHRD-19-0001) of the Indian Institute of Science. AN also acknowledges support from DST-SERB (project number SRG/2020/000153).


\onecolumngrid

\renewcommand\thefigure{S\arabic{figure}}    
\setcounter{figure}{0}   
\renewcommand\theequation{S\arabic{equation}}
\setcounter{equation}{0}

\section{Supplementary Material}

In this supplementary material, we present the following: (I) bandstructure of the antiferromagnetic (AFM) MnIn$_{2}$Te$_{4}$, (II) density of states of ferromagnetic MnIn$_{2}$Te$_{4}$, (III) position of Weyl points in the three-dimensional momentum space, and (IV) anomalous Hall conductivity (AHC).

\section{Bandstructure of antiferromagnetic MnIn$_{2}$Te$_{4}$} 
\begin{figure}[h!]
\includegraphics[scale=0.24]{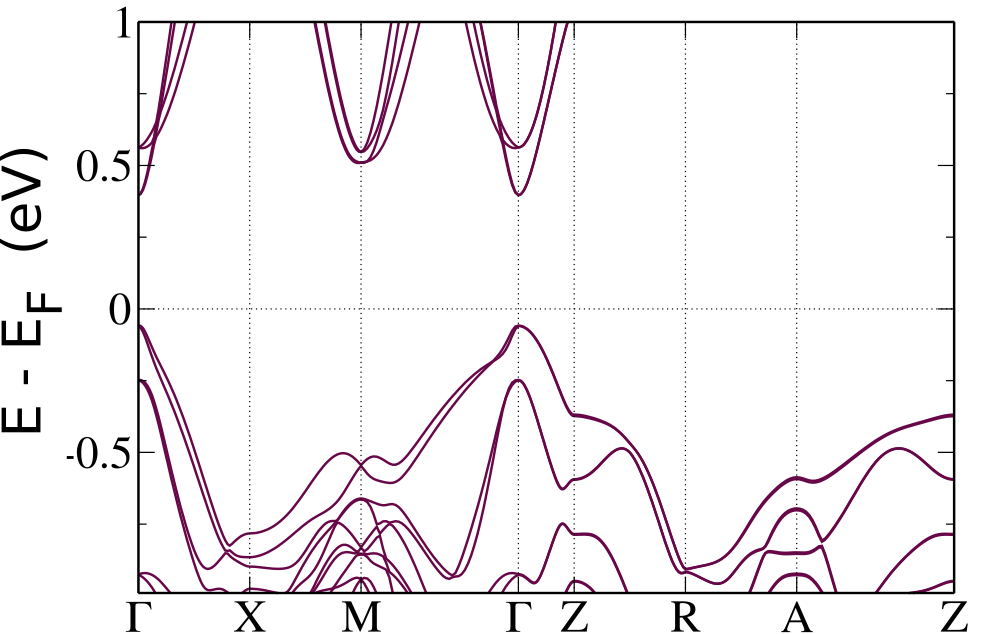}
  \caption{The band structure of antiferromagnetic MnIn$_{2}$Te$_{4}$ at ambient pressure.}
  \label{qe_band}
\end{figure}

Antiferromagnetic MnIn$_{2}$Te$_{4}$ is a trivial magnetic insulator with a direct band gap of $\sim$0.46 eV at $\Gamma$ (see Fig.~\ref{qe_band}, which is almost double that of the ferromagnetic one, discussed in main text). To explore the Weyl physics in antiferromagnetic system, we expect the application of higher pressure. As we noted in the main manuscript, the ferromagnetic structure becomes energetically favorable upon application of small pressure (see main manuscript). 

\section{Density of States}  
\begin{figure}[h!]
\includegraphics[scale=0.25]{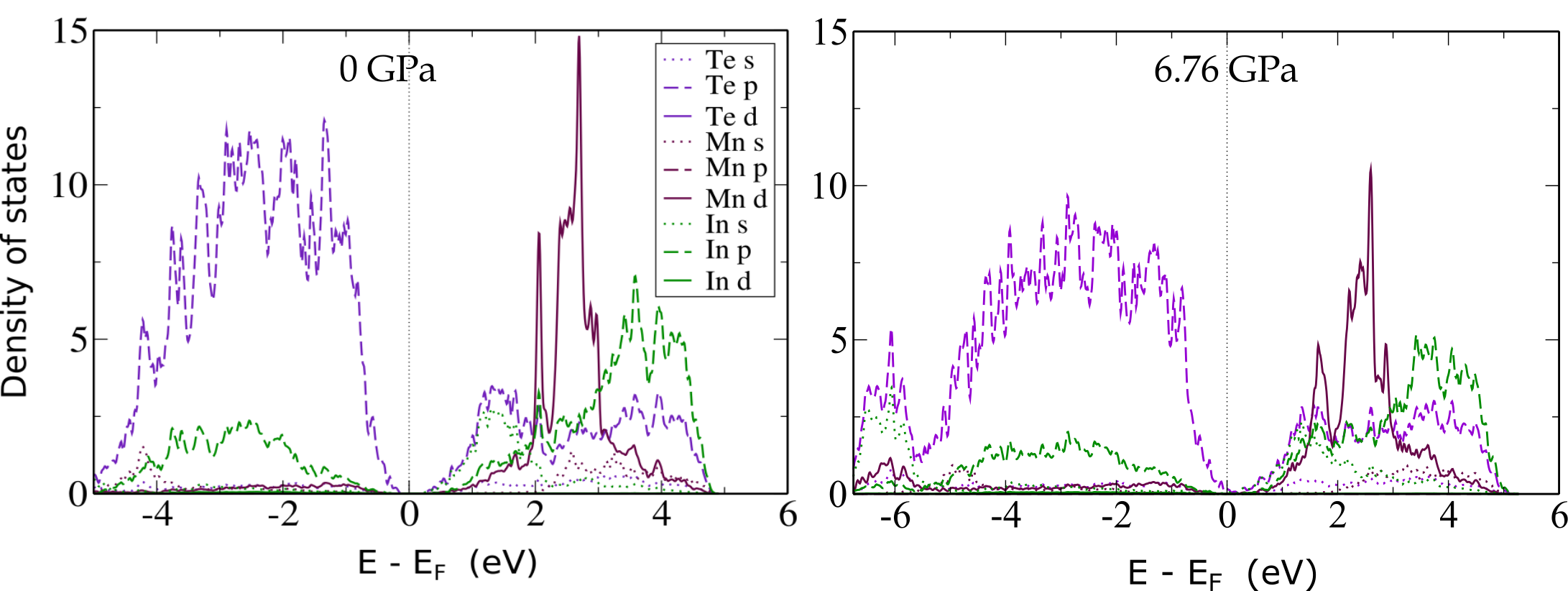}
  \caption{Orbital contribution to the density of states of the ferromagnetic MnIn$_{2}$Te$_{4}$ at ambient (left) and 6.76 GPa (right) pressure.}
  \label{pdos}
\end{figure}

Fig.\,\ref{pdos} shows the partial density of states of ferromagnetic MnIn$_{2}$Te$_{4}$. At non-zero pressures, contribution from the Mn $d$ orbitals increases near the Fermi energy.

\section{Position of Weyl nodes}
\begin{figure}[h!]
\includegraphics[scale=0.22]{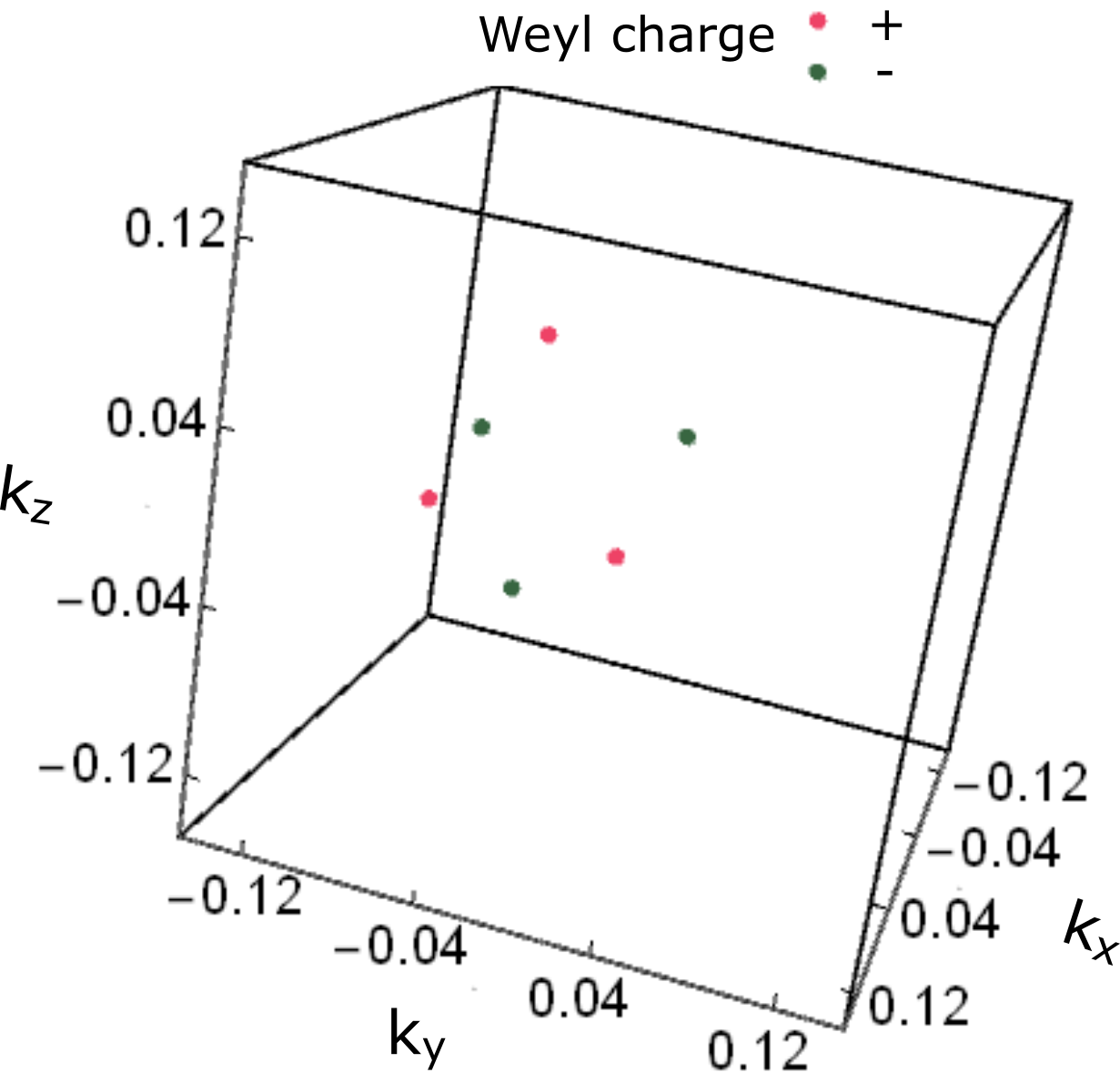}
  \caption{Distribution of Weyl points in the three dimensional momentum space. Red and green dots represent the Weyl nodes with topological charge +1 and -1, respectively, for the ferromagnetic MnIn$_{2}$Te$_{4}$ at 6.76 GPa.}
  \label{nodes}
\end{figure}

Table~\ref{table:bulk_nodes} lists the location of Weyl points in the momentum space, along with their chirality values for the ferromagnetic MnIn$_{2}$Te$_{4}$, at 6.76 GPa. The system contains six Weyl points near the Fermi level -- three points with chirality +1 and the other three with chirality -1. Distribution of these Weyl points in the three-dimensional momentum space is shown in Fig.\,\ref{nodes}.

\begin{table*}[h!]
\caption{Position of bulk nodes at 6.76 GPa.}
\label{table:bulk_nodes}
\begin{tabular}{|p{2.2cm}|p{2.2cm}|p{2.2cm}|p{1.6cm}|}
 \hline
 \hline
 $k_{x}$ (1/\AA) & $k_y$ (1/\AA) & $k_z$ (1/\AA) & Chirality\\
 \hline
 
 0.09629025 & 0.01028182  &  0.00658574  & $+$1\\
  
0.05202581  & $-$0.09052557  & $-$0.00492802 & $+$1\\

$-$0.04290352 &  $-$0.07375086  &  0.05560153 & $+$1\\

$-$0.09621852  & $-$0.00930774  &  0.00493105 & $-$1\\

$-$0.02149350 &  $-$0.09390641 &  0.00941639 & $-$1\\
      
 0.08997877 &  $-$0.03492382 &
$-$0.02169327 & $-$1\\

 \hline
\end{tabular}
\end{table*}

\section{Anomalous Hall Conductivity}
\begin{figure}[h!]
\includegraphics[scale=0.18]{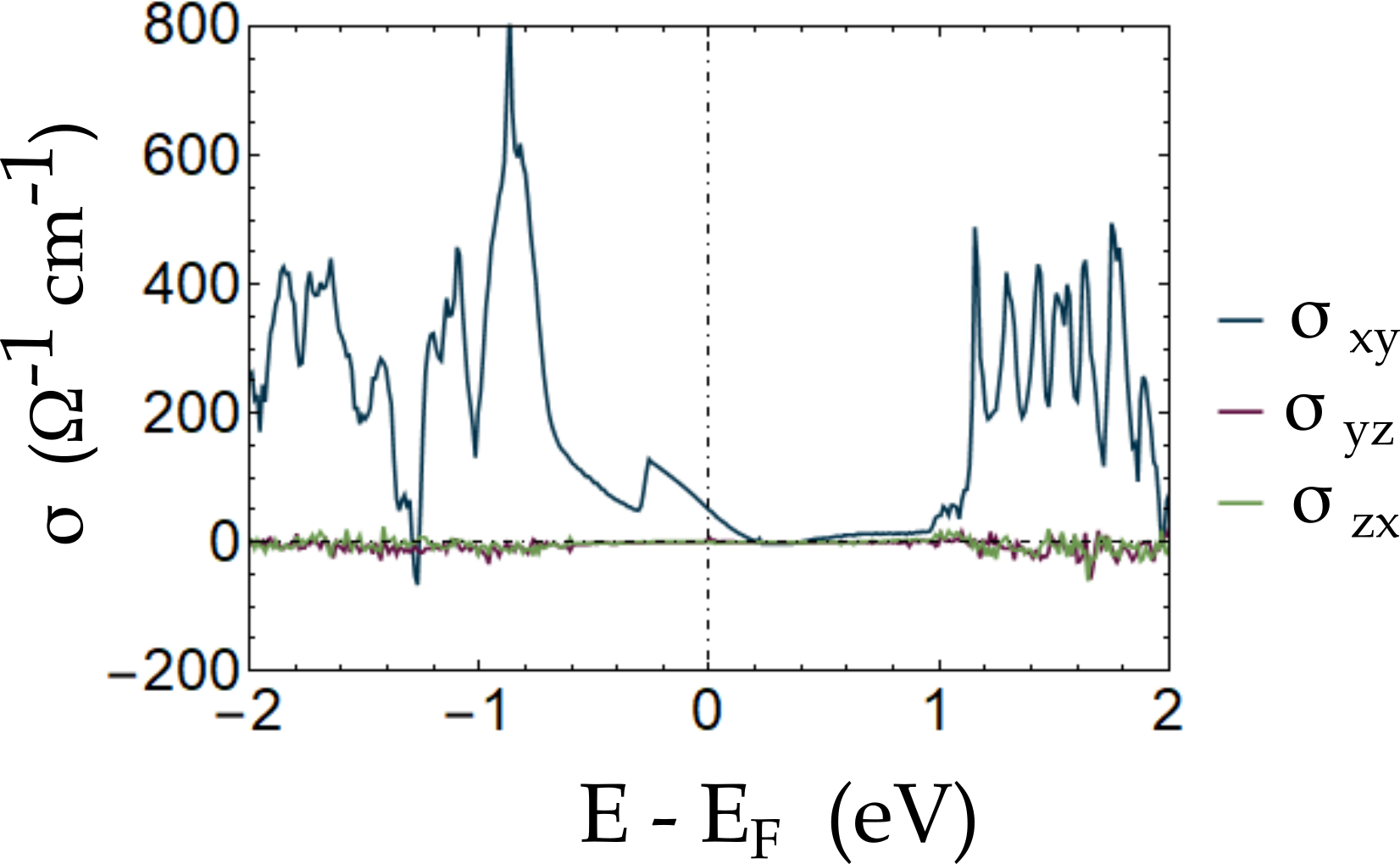}
  \caption{Blue, maroon and light green colours denote the three components of AHC, $\sigma_{xy}$, $\sigma_{yz}$ and $\sigma_{zx}$, respectively. The contributions to the conductivity coming from $\sigma_{yz}$ and $\sigma_{zx}$ are negligible compared to that from $\sigma_{xy}$. }
  \label{AHC_comp}
\end{figure}

In the main text we analyzed the pressure dependence of the AHC signal. Fig.\,\ref{AHC_comp} shows the components of AHC as a function of the Fermi energy. We find that the major contribution to the AHC comes from $\sigma_{xy}$, whereas $\sigma_{yz}$ and $\sigma_{zx}$ contribute negligibly.

\begin{figure}
    \centering
    \includegraphics[scale=0.22]{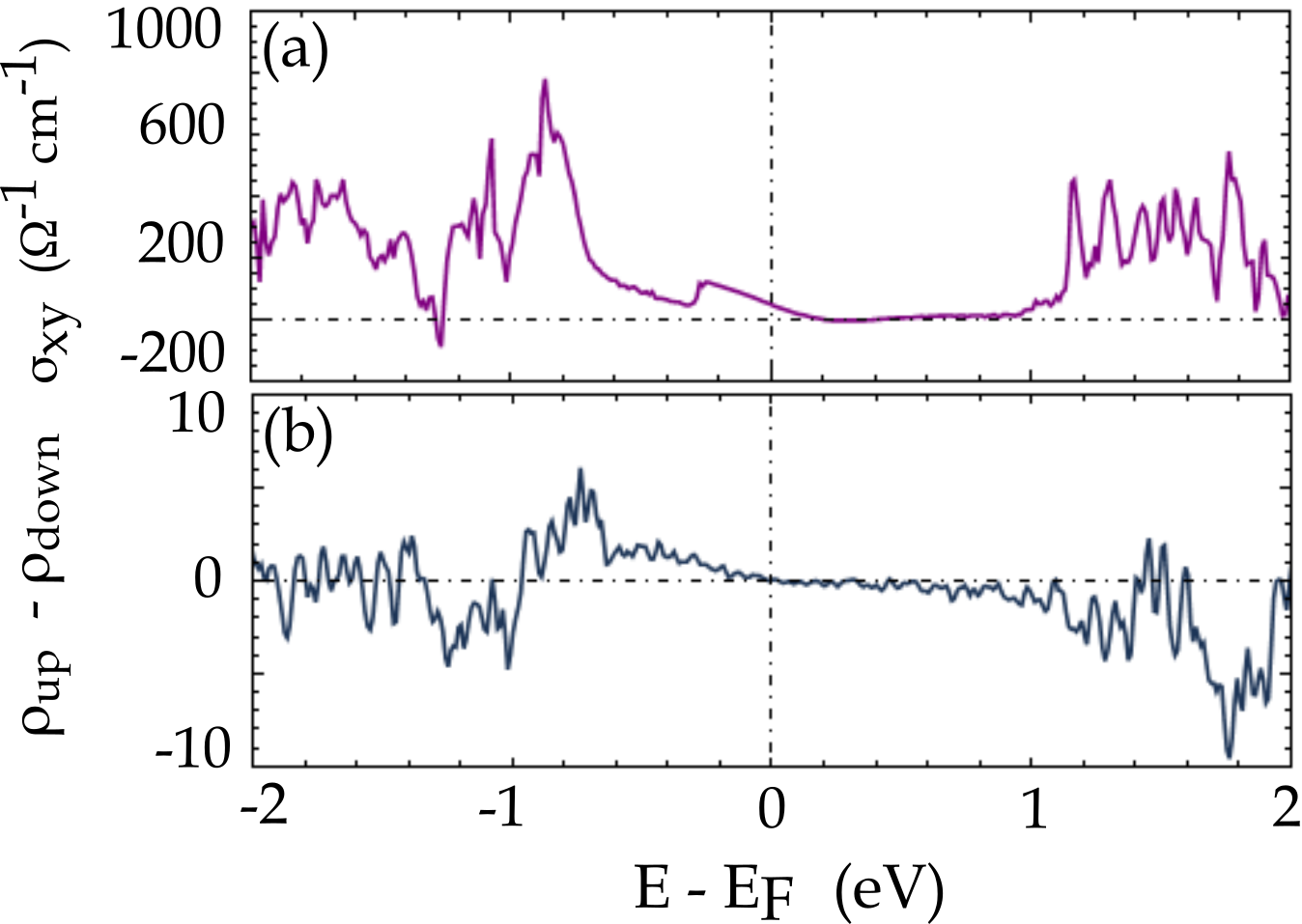}
    \caption{(a) The anomalous Hall conductivity of MnIn$_{2}$Te$_{4}$ at representative 6.76 GPa pressure. The systems shows non-zero ($\sim$50 $\Omega^{-1}$ cm$^{-1}$) AHC at Fermi energy and maintains highest value of $\sim$800 $\Omega^{-1}$ cm$^{-1}$ around 0.8 eV below the Fermi energy. (b) The difference in the density of spin up ($\rho_{up}$) and spin down ($\rho_{down}$) states.}
    \label{AHE}
\end{figure}

Fig.\,\ref{AHE}(a) shows the AHC of the semimetallic phase of FM MnIn$_2$Te$_4$ at a representative hydrostatic pressure value of 6.76 GPa, calculated as a function of Fermi energy.  In the energy range of 0.2 to 0.95 eV, $\sigma_{xy}$ is almost zero. When $E-E_F$ is tuned to zero, i.e., we are at the Fermi energy, we obtain a non-vanishing value of AHC ($\sim$ 50 $\Omega^{-1}$ cm$^{-1}$). It shows a maximum of nearly 800 $\Omega^{-1}$ cm$^{-1}$ around 0.8 eV below the Fermi level. The behaviour of $\sigma_{xy}$, although not directly proportional to the density of available states, matches well with the nature of difference in density of spin up ($\rho_{up}$) and down ($\rho_{down}$) states. Fig.\,\ref{AHE}(b) shows the density difference of up and down spins, $\rho_{up} - \rho_{down} $. We find that the nature of $\sigma_{xy}$ qualitatively follows the density of available states of majority carriers (up or down spin). Close to the Fermi energy, electronic transport is mainly controlled by the Berry curvature around the Weyl nodes. This explains a non-zero $\sigma_{xy}$ at this value in spite of having negligible $\rho_{up}-\rho_{down}$. We note that the AHC can be directly controlled by pressure, since the number of Weyl points is pressure tunable, as we discussed in the main manuscript.


%

\end{document}